\documentclass[pra,twocolumn,showpacs,preprintnumbers,amsmath,amssymb]{revtex4}
\usepackage{graphicx,graphics,color,epsfig,times}
\begin{document}

\title{Implementation of quantum gates based on geometric phases accumulated in
the eigenstates of periodic invariant operators}

\author
{
 L. B. Shao$^{1,2}$, Z. D. Wang$^{2,1*}$, and D. Y. Xing$^{1}$
}
\address{$^{1}$National Laboratory of Solid State of Microstructure and
Department of Physics, \\ Nanjing University, Nanjing 210093, China\\
$^{2}$Department of Physics and Center of Theoretical and
Computational Physics, University of Hong Kong, Pokfulam Road,
Hong Kong, China. }

\begin{abstract}
We propose a new strategy to physically implement  a universal set
of quantum gates based on geometric phases accumulated in the
nondegenerate eigenstates of a designated invariant operator in a
periodic physical system. The system is driven to evolve in such a
way that the dynamical phase shifts of the invariant operator
eigenstates are the same (or {\it mod} $2\pi$) while the
corresponding geometric phases are nontrivial.
 We illustrate how this strategy to work in a simple but typical NMR-type
 qubit system.

\end{abstract}

\pacs{03.67.Lx, 03.65.Vf}

\maketitle

%%%%%%%%%%%%%%%%%%%%%%%%%%%%%%%%%%%%%%%%%%%%%%%%%%%%%%%%%%%%%%%%%%%%%%%%%%%%%%%%

Quantum computation, based on fundamental quantum mechanical
principles such as superposition and entanglement,
%has been believed to be more powerful and efficient in tackling many hard
%problems than its classical counterpart based on Turing
%machines, and thus
may provide a promising perspective to advance modern computational
science~\cite{jozsa,Nielsen}. So far, a lot of substantial efforts
have been dedicated to the field of quantum computation and a number
of significant progresses have been
made~\cite{Nielsen,jic,dva,makhlin}. Nevertheless, quantum
computation is still facing great challenges before it can be put
into practice. As one of them, how to suppress the random errors
during gate operations has been paid much attention for the past
years.

Recently, geometric quantum computation(GQC), expected as an
intrinsical fault-tolerant scheme, was proposed by using
NMR\cite{jaj,wangxb}, superconducting nanocircuits~\cite{falci,zhu},
trapped ions~\cite{duan}, or semiconducting
nanostructures~\cite{solinas}. As is well
known~\cite{berry,simon,a.a}, for an adiabatic(or nonadiabatic)
cyclic evolution, the associated total phase shift consists of both
dynamic and geometric components, where the geometric phase is
interpreted as a holonomy of the Hermitian fiber bundle over the
parameter (projective Hilbert) space.  Since the geometric phase
depends only on the global geometry of the path executed in the
evolution, a set of quantum logical gates related only to the pure
geometric phase shifts in the gate operations are likely to have an
advantage that is insensitive to stochastic operation
errors~\cite{Zanardi,Zhu2006}. A kind of adiabatic GQC based on the
conditional Berry phase was first
proposed~\cite{jaj,falci,duan,solinas}, while the adiabatic
condition may not be satisfied in many realistic cases since it
requires to operate a quantum gate very slowly so that the relevant
instantaneous energy eigenstates follow its Hamiltonian to evolve.
On one hand, faster operation leads to severe distortions in the
expected outcome, while on the other hand, the operation must be
completed within the decoherence time of the system. In order to
overcome this disadvantage, another kind of quantum gates based on
the nonadiabatic geometric phase was suggested~\cite{wangxb,zhu}.
These gates possess likely the virtues of both fast running speed
and intrinsic geometric features of the adiabatic GQC. It is
remarked that a key point in the above conventional GQC schemes is
to avoid or remove the dynamic
phases\cite{jaj,falci,duan,solinas,wangxb,zhu,zhus}, so that only
the geometric phases are accumulated in the whole gate operation.
%Some of them deal with so-called dark states to avoid
%dynamic phase \cite{duan}. Since the dark states are of zero
%energy, the dynamic phases are always zero. Others focus on the
%multiloop schemes that the systems are operated by several special
%closed loops and the total dynamic phase accumulated is zero
%\cite{jaj,zhus}.
Apart from these method, an unconventional GQC was also proposed
to construct quantum gates in several physical
systems~\cite{zhusl,dl}.
%whose dynamic
%phase $\gamma^{d}$ is proportional to geometric phase
%$\gamma^{g}$, that is, $\gamma^{d}=\eta\gamma^{g} (\eta\neq0,-1)$
%with $\eta$ as a proportional constant which is independent on its
%parameters\cite{zhusl,dl}. Thus, the total phase $\gamma$
%accumulated is essentially geometric and this kind of quantum gate
%is also geometric.

In this paper, we propose a new strategy to implement a set of
quantum gates  based on the geometric phases accumulated in the
nondegenerate eigenstates of a periodic invariant operator in a
physical system. Let the system to evolve in such an intriguing way
that the dynamical phase shifts of the invariant operator
eigenstates are the same (or {\it mod} $2\pi$) while the
corresponding geometric phases are nontrivial. In particular, we
illustrate how to realize our scheme in a simple but typical
NMR-type system.  Certainly, the present strategy can also be
applied to other systems.

 Consider a physical system whose Hamiltonian $\hat{H}(t)$ and an
invariant operator $\hat{I}(t)$ are time-dependent and evolve
periodically with periodicity $\tau$~\cite{jeffrey}, i.e.,
\begin{equation} \hat{H}(0)=\hat{H}(\tau),
\qquad \hat{I}(0)=\hat{I}(\tau),
\end{equation}
where the invariant operator $\hat{I}$ is determined by
($\hbar=1$)
\begin{equation} \frac{\partial
\hat{I}}{\partial t}-i[\hat{I},\hat{H}]\equiv 0.
\end{equation}
To realize the geometric quantum gates based on the invariant
operator strategy,  we here focus only on the simple cases in
which all the eigenstates of $\hat{I}$ are nondegenerate. From
Eq.(2) and the eigenvalue equation %of $\hat{I}$
%\begin{equation}
$\hat{I}|n,t\rangle_{I}=\lambda_{n}|n,t\rangle_{I}$ ($n=1,2,...$),
it is straightforward to summarize the following three properties
of $\hat{I}$~\cite{jeffrey,lewis,gao}. (i) All eigenvalues
$\{\lambda_{n}\}$ are time-independent. (ii)The evolving state
$|n,t\rangle_{S}=U(t,0)|n,0\rangle_{I}$ is always the eigenstate
of $\hat{I}(t)$ with the same eigenvalue $\lambda_{n}$, where
$U(t,0)$ is the evolution operator satisfying the Schr\"{o}dinger
equation
%\begin{equation}
$i\frac{\partial}{\partial t}U=\hat{H}U$.
%\end{equation}
Since both $|n,t\rangle_{S}$ and $|n,t\rangle_{I}$ are the
eigenstates of $\hat{I}(t)$ specified by the same eigenvalue
$\lambda_{n}$, there exists a time dependent gauge transformation
between them
\begin{equation}
|n,t\rangle_{S}=e^{i\gamma_{n}(t)}|n,t\rangle_{I}
\end{equation}
with
\begin{equation}
\gamma_{n}(t)=\int_{0}^{t}dt^{'}\langle
n,t^{'}|i\frac{\partial}{\partial
t^{'}}-\hat{H}(t^{'})|n,t^{'}\rangle_{I},
\end{equation}
where the phase $\{\gamma_{n}\}$ is referred to as  ths Lewis
phase\cite{lewis}. (iii) Transitions between the eigenstates
specified by different eigenvalues are impossible, simply because
that the evolution operator represented in the basis
$|n,t\rangle_{I}$ reads %picture of $\hat{I}$ can be written as
\begin{equation}
U_I(t,0)= \left( \begin{array}{ccc}
e^{i\gamma_{1}(t)} & \qquad & \qquad \\
\qquad & e^{i\gamma_{2}(t)} & \qquad \\
\qquad & \qquad & \ddots
\end{array} \right)_I.
\end{equation}
In addition,
%which is a diagonal matrix. The statement outlined above is the
%basic content of the invariant operator.
%\end{equation}
 from  the periodic condition given by Eq.(1), it is straightforward to derive a key relation
$|n,\tau\rangle_{I}=|n,0\rangle_{I}$, which plays an essential
role in the present scheme.

Recently, a theory of geometric phase for invariant operators was
developed\cite{jeffrey}. The corresponding geometric phase is
interpreted as a holonomy inherited from the universal Stiefel
bundle over a Grassmann manifold. For a cyclic evolution of the
eigenstate of $I(t)$, the Lewis phase in Eq.(4) is nothing but the
total phase shift consisting of the geometric phase
$\int_{0}^{\tau}dt^{'}\langle n,t^{'}|i\frac{\partial}{\partial
t^{'}}|n,t^{'}\rangle_{I}$ and the dynamic one
$-\int_{0}^{\tau}dt^{'}\langle
n,t^{'}|\hat{H}(t^{'})|n,t^{'}\rangle_{I}$\cite{a.a,jeffrey,gao}.
%Since the
%evolving state remains as the eigenstate of $\hat{I}$ with the
%same eigenvalue, we consider that a nonadiabatic cyclic evolution
%can be generalized to the adiabatic evolution of the eigenspace of
%invariant operator.
To achieve  quantum gates that depend only on geometric phases, we
need to eliminate the above dynamic phases. We consider a system
whose space bases are the normalized nondegenerate eigenstates of
$\hat{I}(0)$. An arbitrary initial state in the system can be
written as $|\Psi(0)\rangle=\sum_{n}c_{n}|n,0\rangle_{I}$, where
$\{c_{n}\}$ are the expansion coefficients. After operating a
periodic evolution as given in Eq.(1), the final state becomes
\begin{equation}
|\Psi(\tau)\rangle=\sum_{n}c_{n}e^{i\gamma_{n}(\tau)}|n,0\rangle_{I},
\end{equation}
where $\gamma_{n}(\tau)$ is the total phase shift of the state
$|n,0\rangle_{I}$ which consists of the dynamic part
$\gamma_{n}^{d}(\tau)$ and the geometric part
$\gamma_{n}^{g}(\tau)$ \cite{berry,a.a}. Here, we have used the
condition $|n,\tau\rangle_{I}=|n,0\rangle_{I}$. If the accumulated
dynamic phases of $\{|n,0\rangle_{I}\}$ are the same (or {\it mod}
$2\pi$), namely,
\begin{equation}
\gamma_{n}^{d}(\tau)=\gamma_{0}+2K_n\pi,
\end{equation}
where $K_n$ is an integer, the final state $|\Psi(\tau)\rangle$ is
given by
\begin{equation}
|\Psi(\tau)\rangle=e^{i\gamma_{0}}\sum_{n}c_{n}e^{i\gamma_{n}^{g}(\tau)}|n,0\rangle_{I}.
\end{equation}
Note that,  the overall phase shift $\gamma_{0}$ in Eq.(8) is
irrelevant to the designed quantum computation, and thus only the
geometric phase is relevant to the gate operation, which is just a
key idea to construct in principle geometric quantum gates.
% It is
%obvious that our method of eliminating the dynamic phase is more
%involuntary than the conventional ones.
In fact, the present geometric strategy is to operate quantum
gates in such a way that the nondegenerate eigensates of the
invariant operator accumulate the same dynamic phase but with the
nontrivial relative geometric phases.

As a simple but typical example, we now illustrate how to
implement the above generic strategy in an NMR-type qubit system,
noting that the NMR has been a mature technique to simply
simulate/examine quantum information processing schemes.
Certainly, the present scheme is also applicable in principle to
other quantum systems that evolve nonadiabatically. Consider an
NMR-type spin-1/2 system, subject to a rotating magnetic field
%applied in the direction of $\hat{x}$, $\hat{y}$ and $\hat{z}$ is
given by
\begin{equation}
{\vec B}(t)=(-B_{2}\cos\omega t,-B_{2}\sin\omega t,-B_{1}),
\end{equation}
 where $B_{1}$ and $B_{2}$
 are respectively the amplitudes of $z$ and $xy$-plane components  of the
 field. The corresponding qubit Hamiltonian can be written as
\begin{equation}
\hat{H}(t)=-\frac{1}{2}\mu_{B}{\vec \sigma} \cdot {\vec
B}=\frac{1}{2}\omega_{1}\sigma_{z}+\frac{1}{2}\omega_{2}\left(
\begin{array}{ccc} 0 & e^{-i\omega t} \\ e^{i\omega t} & 0 \end{array}
\right),
\end{equation}
where $\sigma_{x,y,z}$ are the Pauli matrices, $\omega_{1}=\mu_{B}
B_{1}$ and $\omega_{2}=\mu_{B} B_{2}$ with $\mu_{B}$ as the Bohr
magneton. The corresponding evolution operator  is~\cite{LMK}
\begin{equation}
U(t,0)=e^{-i\omega t\sigma_{z}/2}e^{-iH_{0}t},
\end{equation}
where $\hat{H}_{0}=\hat{H}(0)-\omega\sigma_{z}/2$ is just the
Hamiltonian denoted  in the rotating framework. An invariant
operator satisfying Eq.(2) is then found to be
\begin{equation}
\hat{I}=\frac{1}{2}(\omega_{1}-\omega)\sigma_{z}+\frac{1}{2}\omega_{2}\left(
\begin{array}{ccc} 0 & e^{-i\omega t} \\ e^{i\omega t} & 0 \end{array}
\right).
\end{equation}
Obviously, $\hat{I}(0)=\hat{H}_{0}$, and the invariant operator
follows the Hamiltonian to evolve cyclically with periodicity
$\tau=2\pi/\omega$. The eigenvalues of $\hat{I}$ are evaluated to be
$\pm \lambda/2$ with
$\lambda=\sqrt{\omega_{2}^{2}+(\omega_{1}-\omega)^{2}}$, and the two
corresponding eigenstates are derived as
$|\frac{1}{2}\lambda,t\rangle_{I}=\cos\frac{\chi}{2}|\uparrow\rangle+e^{i\omega
t}\sin\frac{\chi}{2}|\downarrow\rangle$ and
$|-\frac{1}{2}\lambda,t\rangle_{I}=-\sin\frac{\chi}{2}|\uparrow\rangle+e^{i\omega
t}\cos\frac{\chi}{2}|\downarrow\rangle$, where
$\chi=2\arctan\frac{\lambda+\omega-\omega_{1}}{\omega_{2}}$,
$|\uparrow\rangle$ and $|\downarrow\rangle$ are the two eigenstates
of $\sigma_{z}$. Since $\hat{I}(0)=\hat{H}_{0}$, $|\pm
\frac{1}{2}\lambda,0\rangle_{I}$ are also the eigenstates of
$H_{0}$.  For a cyclic evolution and in the basis $|\pm
\frac{1}{2}\lambda,0\rangle_{I}$, the evolution operator
$U_{I}(\tau)$ can be simply written as
%$|\pm\frac{1}{2}\lambda,0\rangle_{I}$ as
\begin{equation}
U_{I}(\tau,0) = \left( \begin{array}{cc}
e^{i\pi(1-\lambda/\omega)} & 0 \\
0 & e^{i\pi(1+\lambda/\omega)}
\end{array} \right)_I.
\end{equation}
Note that, if we choose the computation basis as
$|\uparrow\rangle$ and $|\downarrow\rangle$,
%with $|\frac{1}{2}\lambda,0\rangle_{I}=\cos\frac{\chi}{2}|\uparrow\rangle+\sin\frac{\chi}{2}|\downarrow\rangle$
%and $|-\frac{1}{2}\lambda,0\rangle_{I}=-\sin\frac{\chi}{2}|\uparrow\rangle+\cos\frac{\chi}{2}|\downarrow\rangle$,
the unitary transformation ${\tilde U}(\tau, 0)$ between the input
and output states can also be spelt out explicitly~\cite{zhangxd}
%\begin{equation}
 $${\tilde U}
 %(\gamma_+, \chi)
 =\left (
\begin{array}{ll}
e^{i\gamma}\cos^2\frac{\chi}{2}+e^{-i\gamma}\sin^2\frac{\chi}{2}
& i\sin \chi \sin\gamma  \\
i \sin\chi \sin\gamma &
e^{i\gamma}\sin^2\frac{\chi}{2}+e^{-i\gamma}\cos^2\frac{\chi}{2}
\end{array}
\right ), $$
%\end{equation}
where $\gamma=\gamma_+=\pi(1-\lambda/\omega)$ is the total phase
shift of $|\frac{1}{2}\lambda,0\rangle_{I}$ in one cyclic
evolution.

 Similarly, the total phases shift of $|-\frac{1}{2}\lambda,0\rangle_{I}$ is
 expressed as
 $\gamma_-=\pi(1+\lambda/\omega)$ in the cyclic evolution.  The
corresponding dynamic phases are derived to be
\begin{equation}
\gamma_{\pm}^{d}=\mp\pi\frac{\sqrt{\omega_{1}^{2}+\omega_{2}^{2}}}{\omega}\cos(\chi-\theta),
\end{equation}
with $\theta=\arctan(\omega_{2}/\omega_{1})$. Here
% From Eq.(), it is seen that since ,
$(\chi-\theta)$ is just the angle between the magnet field and the
state vector in the Bloch sphere as the eigenstate of $\hat{I}$
rotates with $\hat{H}$. The geometric phases of
$|\pm\frac{1}{2}\lambda,0\rangle_{I}$ are found to be
\begin{equation}
\gamma_{\pm}^{g}=\pi(1\pm\cos\chi).
\end{equation}
Using Eq.(7)) to eliminate the dynamic phases in the gate operation,
we are able to derive a relation for three parameters $\omega_{1}$,
$\omega_{2}$ and $\omega$
\begin{equation}
\frac{(\lambda+\omega-\omega_{1})
(\omega_{1}^{2}-\omega\omega_{1}+\omega_{2}^{2})}{\omega_2^{2}+(\lambda+\omega-\omega_{1})^{2}}
=\frac{K\omega}{2},
\end{equation}
where $K$ is an integer. In the simplest case of $K=0$, we have
$\omega_{1}^{2}+\omega_{2}^{2}=\omega\omega_{1}$. The geometric
phases are simply given by
\begin{equation}
\gamma_{\pm}^{g}=\pi(1\mp\frac{\lambda}{\omega})=\pi(1\pm\cos\chi).
\end{equation}

Comparing with the existing GQC schemes~\cite{zhu,zhangxd}, the
present strategy is simpler and more operable. Also interestingly,
in the adiabatic evolution, i.e., $\omega\ll\omega_{1},\omega_{2}$,
one has $\partial \hat{I}/\partial t\approx0$, so that
$[\hat{I},\hat{H}]\approx 0$ and $\hat{I}\approx \hat{H}$ in the
present example. In this case, the two eigenstates of $\hat{I}$ are
also the instantaneous eigenstates of $\hat{H}$ and the conventional
Berry phase is recovered~\cite{jeffrey}. Moreover, under the
adiabatic approximation, a set of adiabatic Abelian geometric gates
can be constructed more rigorously using the present theory plus a
two-loop gate operation that can simply eliminate the dynamic phase.

At this stage, we elaborate how to realize a universal two-qubit
quantum gate, namely, to construct a controlled-$U$ gate given by
the following unitary transformation
\begin{equation}
U_{c}(T)=\left( \begin{array}{cc} E & O \\
O & U
\end{array} \right),
\end{equation}
where $E$ and $O$ represent respectively the $2\times 2$ unitary
and zero matrixes , and $T$ is the operation periodicity of the
gate. For simplicity but without loss of generality, we consider
that the two spin-1/2 systems are coupled by an ordinary
$-J\sigma_{1z}\sigma_{2z}/2$ term with the coupling strength $J$.
We also set $B_{1}=0$ for each single qubit and let  the two
resonant magnetic fields are applied only  on the first qubit. The
total Hamiltonian for this two-qubit system reads
\begin{equation}
\hat{H}_t=-\frac{1}{2}J\sigma_{1z}\sigma_{2z}+\frac{1}{2}\omega_{0}(\sigma_{1x}\cos\omega
t+\sigma_{1y}\sin \omega t)
\end{equation}
where $\sigma_{1}$ and $\sigma_{2}$ are the Pauli matrix for
qubits $1$ and $2$, respectively. In the representation of
$|\uparrow\uparrow\rangle,|\downarrow\uparrow\rangle,|\uparrow\downarrow\rangle$
and $|\downarrow\downarrow\rangle$, $\hat{H}_t$ can be decomposed
into a direct product of the two single-qubit Hamiltonians as
\begin{equation}
\hat{H}_t=\hat{H}_{1}\otimes\hat{H}_{2},
\end{equation}
where $\hat{H}_{1}$ and $\hat{H}_{2}$ are written as
\begin{displaymath}
\hat{H}_{1}=\frac{1}{2}\left( \begin{array}{ccc} -J &
\omega_{0}e^{-i\omega t} \\ \omega_{0}e^{i\omega t} & J\end{array}
\right)
\end{displaymath}
and
\begin{displaymath}
\hat{H}_{2}=\frac{1}{2}\left( \begin{array}{ccc} J &
\omega_{0}e^{-i\omega t} \\ \omega_{0}e^{i\omega t} &
-J\end{array} \right).
\end{displaymath}
Clearly, $\hat{H}_{1}$ corresponds to the subspace spanned by bases
$|\uparrow\uparrow\rangle$ and $|\downarrow\uparrow\rangle$, while
$\hat{H}_{2}$ is in the subspace spanned by  bases
$|\uparrow\downarrow \rangle$ and $|\downarrow\downarrow\rangle$.
These two subspace are orthogonal. The invariant operator in Eq.(2)
is found to be
\begin{equation}
I_t=I_{1}\otimes I_{2},
\end{equation}
where
\begin{displaymath}
I_{1}=\frac{1}{2}\left( \begin{array}{ccc} -J-\omega &
\omega_{0}e^{-i\omega t} \\ \omega_{0}e^{i\omega t} & J+\omega
\end{array} \right)
\end{displaymath}
and
\begin{displaymath}
I_{2}=\frac{1}{2}\left( \begin{array}{ccc} J-\omega &
\omega_{0}e^{-i\omega t} \\ \omega_{0}e^{i\omega t} &
-J+\omega\end{array} \right).
\end{displaymath}
The eigenvalues are $\pm\lambda_{1}/2$ and $\pm\lambda_{2}/2$,
where $\lambda_{1}=\sqrt{\omega_{0}^{2}+(J+\omega)^{2}}$ and
$\lambda_{2}=\sqrt{\omega_{0}^{2}+(J-\omega)^{2}}$, respectively.
The corresponding eigenstates of $I$ are
$|\frac{1}{2}\lambda_{1},t\rangle=\cos\frac{\chi_{1}}{2}|\uparrow\uparrow\rangle+e^{i\omega
t}\sin\frac{\chi_{1}}{2}|\uparrow\downarrow\rangle$,
$|-\frac{1}{2}\lambda_{1},t\rangle=-\sin\frac{\chi_{1}}{2}|\uparrow\uparrow\rangle+e^{i\omega
t}\cos\frac{\chi_{1}}{2}|\uparrow\downarrow\rangle$,
$|\frac{1}{2}\lambda_{2},t\rangle=\cos\frac{\chi_{2}}{2}|\uparrow\downarrow\rangle+e^{i\omega
t}\sin\frac{\chi_{2}}{2}|\downarrow\downarrow\rangle$ and
$|-\frac{1}{2}\lambda_{2},t\rangle=-\sin\frac{\chi_{2}}{2}|\uparrow\downarrow\rangle+e^{i\omega
t}\cos\frac{\chi_{2}}{2}|\downarrow\downarrow\rangle$, where
$\chi_{1}=2\arctan\frac{\lambda_{1}+\omega+J}{\omega_{0}}$ and
$\chi_{2}=2\arctan\frac{\lambda_{2}+\omega-J}{\omega_{0}}$. For a
cyclic evolution, the evolution operator of $\hat{H}_t$ in the
representation of $|\pm\frac{1}{2}\lambda_{1},0\rangle$ and
$|\pm\frac{1}{2}\lambda_{2},0\rangle$ can be written as
\begin{equation}
U_I(\tau)=U_{1}(\tau)\otimes U_{2}(\tau),
\end{equation}
where
\begin{displaymath} U_{1}(\tau)=\left( \begin{array}{cc}
e^{i\pi(1-\lambda_{1}/\omega)} & 0 \\
0 & e^{i\pi(1+\lambda_{1}/\omega)}
\end{array} \right)_I
\end{displaymath}
and
\begin{displaymath}
U_{2}(\tau)=\left( \begin{array}{cc}
e^{i\pi(1-\lambda_{2}/\omega)} & 0 \\
0 & e^{i\pi(1+\lambda_{2}/\omega)}
\end{array} \right)_I.
\end{displaymath}
By a close inspection on Eq.(22) and considering that
$\lambda_{1}\neq\lambda_{2}$, we find that once  a multi-cycle
evolution $T=m\tau$ is operated, one may be able to achieve a
two-qubut gate given by Eq.(18). For such an operation, the unitary
transformation is given by
\begin{equation}
U_I(T)=U_I(m\tau)=U_I(\tau)^{m}.
\end{equation}
Theoretically, for a rational $(\lambda_{1}/\omega)$, one may
assert the relation
\begin{equation}
m\pi(1+\lambda_{1}/\omega)=2N\pi,
\end{equation}
to be satisfied, where $N$ is an integer. This condition makes
$U_{1}$ to be a unitary matrix, namely,
\begin{equation}
U_I(T)=\left( \begin{array}{cc} E & O \\
O & U_{2}(m\tau)
\end{array} \right)_I.
\end{equation}
Correspondingly, in the bases $(|\uparrow\uparrow\rangle$,
$|\downarrow\uparrow\rangle$, $|\uparrow\downarrow \rangle$,
$|\downarrow\downarrow\rangle)$, we  have also a controlled-U gate
in the form
\begin{equation}
{\tilde U}_{c}(T)=\left( \begin{array}{cc} E & O \\
O & {\tilde U}_{2}(m\tau)
\end{array} \right).
\end{equation}
 Under this
multi-cycle operation, the dynamic phase of $U_2$ is
$\gamma_{\pm}^{d}(m\tau)=m\gamma_{\pm}^{d}(\tau)$. Thus we choose
the appropriate parameters to ensure Eq. (16), i.e.,
\begin{equation}
\frac{(\lambda_{2}+\omega-J)(J^{2}-J\omega+\omega_{0}^{2})}{\omega_2^{2}+
(\lambda_{2}+\omega-J)^{2}}=\frac{K\omega}{2m}.
\end{equation}
When qubit 2 is down, the geometric phase for $K=0$ is
\begin{equation}
\gamma_{2\pm}^{g}=m\pi(1\mp\frac{\lambda_{2}}{\omega}).
\end{equation}
As a result, a universal geometric quantum gate is realized. For
instance, if we set $K=0$, $N=4$, $J=\frac{16}{27}\omega$, and
$\omega_{0}=\frac{4\sqrt{11}}{27}\omega$, we could have $m=3$. The
geometric phases are thus
$\gamma^g_{2\pm}=\pi(1\mp\sqrt{\frac{11}{3}})$.
%One may oppugn that
%this scheme of operation will induce errors and decrease the running
%speed. By contrast, if we have detected or predicted some
%perturbation in the previous cyclic evolution , we can correct
%induced errors in the sequent cyclic evolutions. Fortunately, both
%Eq.(26) and Eq.(28) are determined by the ratios of the parameters,
%so that we can increase the value of $\omega$ to accelerate the
%running speed.

In summary, we have proposed a new strategy to implement a set of
quantum gates  based on the geometric phases accumulated in the
nondegenerate eigenstates of an invariant operator in a periodic
physical system.  An intriguing way is presented to eliminate the
dynamical phase shifts in the designated gate operation. In
addition, we have also illustrated how to implement our scheme in a
simple but typical NMR-type system, while the present strategy may
also be feasible in other systems.

\begin{acknowledgments}
We thank J. C. Y. Teo for helpful discussions. This work was
supported by the National Natural Science Foundation of China
(10429401),
%the Ministry of Science and Technology of China through
%the 973 program (2006CB601007),
the RGC of Hong Kong (HKU 7045/05), and the URC fund of Hong Kong.
\end{acknowledgments}
* E-mail: zwang@hkucc.hku.hk
%%%%%%%%%%%%%%%%%%%%%%%%%%%%%%%%%%%%%%%%%%%%%%%%%%%%%%%%%%%%%%%%%%%%%%%%%%%%%%%%%%%%%%%%%%%%%%%%

%%%%%%%%%%%%%%%%%%%%%%%%%%%%%%%%%%%%%%%%%%%%%%%%%%%%%%%%%%%%%%%%%%%%%%%%%%%%%%%%%%%%%%%%%%%%%%%%%

\end{document}